\documentclass{ifacconf}
\usepackage{graphicx}      
\usepackage{natbib}        
\graphicspath{ {./figures/} }
\usepackage{subcaption}
\usepackage{amssymb}
\usepackage{lineno}
\usepackage{mathtools}
\usepackage{comment}
\newtheorem{theorem}{Theorem}

\newtheorem{assumption}{Assumption}

\usepackage{tikz}
\usetikzlibrary{arrows, positioning, calc}
\usepackage{environ}

\makeatletter
\newsavebox{\measure@tikzpicture}
\NewEnviron{scaletikzpicturetowidth}[1]{%
	\def\tikz@width{#1}%
	\def\tikzscale{1}\begin{lrbox}{\measure@tikzpicture}%
		\BODY
	\end{lrbox}%
	\pgfmathparse{#1/\wd\measure@tikzpicture}%
	\edef\tikzscale{\pgfmathresult}%
	\BODY
    }
\begin{document}
\begin{frontmatter}

\title{Higher-Order Harmonics Reduction in Reset-Based Control Systems: Application to Precision Positioning Systems} 

\thanks[footnoteinfo]{This work is supported by ASMPT, 6641 TL, The Netherlands.}

\author[First]{S. Ali Hosseini} 
\author[Second]{Nima Karbasizadeh} 
\author[First]{S. Hassan HosseiniNia}

\address[First]{Department of Precision and Microsystems
Engineering, Delft University of Technology, 2628 CD Delft, The Netherlands (e-mail: (s.a.hosseini,
s.h.hosseinniakani)@tudelft.nl).}
\address[Second]{ASML, 5504 DR Eindhoven, The Netherlands (e-mail: nima.karbasi@asml.com)}

\begin{abstract}                
To address the limitations imposed by Bode's gain-phase relationship in linear controllers, a reset-based filter called the Constant in gain- Lead in phase (CgLp) filter has been introduced. This filter consists of a reset element and a linear lead filter. However, the sequencing of these two components has been a topic of debate. Positioning the lead filter before the reset element in the loop leads to noise amplification in the reset signal, whereas placing the lead filter after the reset element results in the magnification of higher-order harmonics. This study introduces a tunable lead CgLp structure in which the lead filter is divided into two segments, enabling a balance between noise reduction and higher-order harmonics mitigation. Additionally, a filtering technique is proposed, employing a target-frequency-based approach to mitigate nonlinearity in reset control systems in the presence of noise. The effectiveness of the proposed methods in reducing nonlinearity is demonstrated through both frequency domain and time-domain analyses using a simulated precision positioning system as a case study.
\end{abstract}
\begin{keyword}
Nonlinear control, Reset control, Frequency domain analysis, Linear control limitations, Describing functions, Motion control.
\end{keyword}

\end{frontmatter}
\section{Introduction}
Reset elements are nonlinear filters used to overcome the fundamental performance limitations of linear time-invariant (LTI) control systems \cite{zhao2019overcoming}. The concept of reset control was initially introduced in \cite{clegg1958nonlinear} as a nonlinear integrator, later known as the Clegg integrator (CI).
It demonstrated promising behavior in overcoming the limitations inherent in linear feedback control caused by Bode’s gain-phase relationship \cite{guo2009frequency}.
Over time, more advanced reset components were created, including the generalized first-order reset element (GFORE) \cite{guo2009frequency} and the second-order reset element (SORE) \cite{hazeleger2016second}.

In \cite{guo2009frequency}, the reset element is presented in the frequency domain using the describing function method. In \cite{saikumar2021loop}, the extension of the frequency domain tool ‘Higher Order Sinusoidal-Input Describing Functions (HOSIDFs)’ for reset controllers is introduced, enabling deeper open-loop analysis. Additionally, in \cite{saikumar2021loop}, a method for translating open-loop behavior to closed-loop in the frequency domain is proposed, which corresponds to the HOSIDF-based sensitivity functions for reset control systems (RCS).

In \cite{saikumar2019constant}, the Constant in gain, Lead in phase (CgLp) element, a reset-based filter, is introduced as a remedy for Bode’s gain-phase relationship. CgLp element is able to produce a positive phase while having constant gain at its first-order Sinusoidal Input Describing Function (SIDF). This filter consists of a GFORE and a lead element. In \cite{cai2020optimal}, it is shown that having a lead action before the reset element results in a smaller magnitude of HOSIDFs. Thus, employing a Lead-GFORE sequence for the CgLp can reduce nonlinearity. However, in the presence of noise in the system, a lead action before the reset element amplifies the noise in the reset action signal, leading to unwanted resets. Therefore, in this study, by dividing a lead filter into two parts, we introduce:

\begin{itemize} \item Tunable Lead CgLp, a method to compromise the effects of noise and HOSIDFs on RCSs performance. \end{itemize}
This method can reduce nonlinearity to the extent permitted by the noise.

In \cite{karbasizadeh2022continuous}, a pre- and post-filtering technique using lead and lag filters is introduced to reduce HOSIDFs and improve transient response. However, its effectiveness is limited by noise, as the lead filter precedes the reset element and operates across a broad frequency range despite HOSIDFs being confined to a specific band. To address nonlinearity within a targeted frequency range, this study proposes the following method:

\begin{itemize} \item Notch-based filtering of the CgLp element, a direct closed-loop approach for shaping nonlinearity in the presence of noise. \end{itemize}

In this paper, first, we introduce the reset element in both the time and frequency domains in Section \ref{sec: priliminaries}. Since our objective is to shape the nonlinearity using the closed-loop frequency domain characteristics of the RCS, Section \ref{sec: case study} introduces the case study along with the linear and reset controllers utilized in this paper. Then in Section \ref{sec: shaping}, the methodology of the Tunable Lead CgLp and the Filtered CgLp are presented and in Section \ref{Sec: validation} we validated the results in both time and frequency domains.
\section{Preliminaries}
\label{sec: priliminaries}
\subsection{System description}
Consider a closed-loop control system as Fig. \ref{fig:Block diagram cl}, where $G$ is the plant, $C_1$ and $C_2$ are LTI filters and $\mathcal{R}$ is the reset element describes as follows
\begin{equation} 
		\mathcal{R} : \begin{cases}
            \dot{x}_r(t)=A_rx_r(t)+B_re_r(t),  \quad \quad \left(x_r(t),e_r(t)\right)\notin\mathcal{F},\\
            x_r(t^+)=A_\rho x_r(t), \qquad \qquad \qquad\left(x_r(t),e_r(t)\right)\in\mathcal{F},\\
            u_r(t)=C_rx_r(t)+D_re_r(t),
		\end{cases}
  \label{eq.SS reset}
	\end{equation}
 with
 \begin{equation}
    \label{reset surface}
\mathcal{F}:=\{e_r=0\wedge(A_\rho-I)x_r(t)\neq0\},
\end{equation}
where $A_r\in\mathbb{R}^{n_r\, . \, n_r}$, $B_r\in\mathbb{R}^{n_r\, . \, 1}$, $C_r\in\mathbb{R}^{1\, . \, n_r}$, and $D_r\in\mathbb{R}$ represent the state-space matrices of the reset element, and the reset value matrix is denoted by $A_\rho=\text{diag}(\gamma_1,...,\gamma_{n_r})$. $x_r(t)\in\mathbb{R}^{n_r\, . \, 1}$ is the state of the reset element, $x_r(t^{+})\in\mathbb{R}^{n_r\, . \, 1}$ is the after-reset state, and $e_r(t)\in\mathbb{R}$ and $u_r(t)\in\mathbb{R}$ represent the input and output of the reset element, respectively. By defining the base linear system of the reset element as the case when there is no reset, its transfer function is calculated as follows:
\begin{equation}
\label{eq RCS bls}
R(s) = C_r(s - A_r)^{-1}B_r + D_r,
\end{equation}
where $s \in\mathbb{C}$ is the Laplace variable.

\usetikzlibrary {arrows.meta}
\tikzstyle{block} = [draw,thick, fill=white, rectangle, minimum height=2em, minimum width=2.5em, anchor=center]
\tikzstyle{sum} = [draw, fill=white, circle, minimum height=0.6em, minimum width=0.6em, anchor=center, inner sep=0pt]
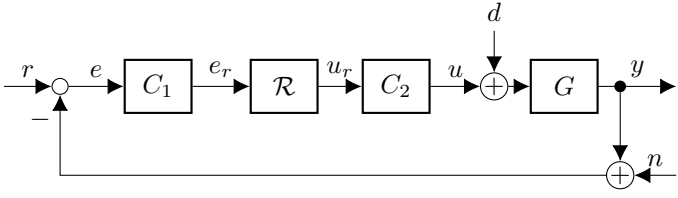
\begin{figure}[!t]
	\centering
	\begin{scaletikzpicturetowidth}{\linewidth}
		\begin{tikzpicture}[scale=\tikzscale]
			\node[coordinate](input) at (0,0) {};
			\node[sum] (sum1) at (1,0) {};
			\node[sum] (sum3) at (8.75,0) {+};
			\node[sum] (sum4) at (11,-1.6) {+};
			\node[sum, fill=black, minimum size=0.4em] (dot2) at (11,0) {};		
			
			\node[block] (lead) at (5.0,0) {$\mathcal{R}$};
			\node[block] (controller) at (7.0,0) {$C_2$};
			\node[block] (fo-higs) at (2.75,0) {$C_1$};
			\node[block] (system) at (10,0) {$G$};
			\node[coordinate](output) at (12,0) {};
			\node[coordinate](di-input) at (8.75,1) {};
			\node[coordinate](n-input) at (12,-1.6) {};
			\draw[arrows = {-Latex[width=6pt, length=6pt]}] (input)  -- node[above]{$r$} (sum1);
			\draw[arrows = {-Latex[width=6pt, length=6pt]}] (di-input)node[above]{$d$}  --  (sum3);
			\draw[arrows = {-Latex[width=6pt, length=6pt]}] (n-input)  -- node[above]{$n$} (sum4);
			\draw[arrows = {-Latex[width=6pt, length=6pt]}] (sum1)   --node[above]{$e$}  (fo-higs);
			\draw[arrows = {-Latex[width=6pt, length=6pt]}] (fo-higs) --node [above]{$e_r$} (lead);
			\draw[arrows = {-Latex[width=6pt, length=6pt]}] (lead)  --node[above] {$u_r$} (controller);
			\draw[arrows = {-Latex[width=6pt, length=6pt]}] (controller)  --node[above] {$u$} (sum3);
			\draw[arrows = {-Latex[width=6pt, length=6pt]}] (sum3) -- (system);
			\draw[arrows = {-Latex[width=6pt, length=6pt]}] (system)  -- node[above]{$y$} (output);
			\draw[arrows = {-Latex[width=6pt, length=6pt]}] (dot2) -- (sum4);
			\draw[arrows = {-Latex[width=6pt, length=6pt]}] (sum4) -| node[pos=0.85,left]{$-$} (sum1);
			
		\end{tikzpicture}
	\end{scaletikzpicturetowidth}
	\caption{Block diagram of the closed-loop system.}
	\label{fig:Block diagram cl}
\end{figure}
\subsection{Open-loop and Closed-loop Frequency Domain Analysis}
\begin{theorem}
    \label{theorem reset hosidf}
    \cite[Theorem 3.1]{saikumar2021loop},
    Considering $e_r(t)=\sin{(\omega t)}$ as the input of the reset element, then for its output, we have 
    \begin{equation}
        u_r(t)=\sum^{\infty}_{n=1}{|H_n(\omega)|\sin{\left(n\omega t+\angle{H_n(\omega)}\right)}},
    \end{equation}
     where $H_n(\omega)$ is the HOSIDF of the reset element in \eqref{eq.SS reset} as follows
    \begin{equation} 
    \label{eq reset HOSIDF}
    \resizebox{1\hsize}{!}{
		$H_n(\omega) = \begin{cases}
            C_r\left(j\omega I-A_r\right)^{-1}\left(I+j\Theta_D(\omega)\right)B_r+D_r,  \quad \quad \mathrm{for}\;\; n=1,\\
            C_r\left(jn\omega I-A_r\right)^{-1}j\Theta_D(\omega)B_r,  \qquad \qquad \hspace{9mm} \mathrm{for}\;\;\mathrm{odd}\;\; n\geq2,\\
            0, \hspace{66mm} \mathrm{for}\;\;\mathrm{even}\;\; n\geq2,
		\end{cases}$
  }
	\end{equation}
 with
 \begin{equation}
 \begin{split}
&\Lambda(\omega)=\omega^2 I+A_{r}^2, \quad \Delta(\omega)=I+e^{\frac{\pi}{\omega}A_r},  \\
&\Delta_r(\omega)=I+A_\rho e^{\frac{\pi}{\omega}A_r}, \quad \Gamma_r(\omega)=\Delta_r^{-1}(\omega)A_\rho\Delta(\omega)\Lambda^{-1}(\omega),\\ 
     &\Theta_\mathrm{D}(\omega)=-\frac{2\omega^2}{\pi}\Delta(\omega)\left[\Gamma_r(\omega)-\Lambda^{-1}(\omega)\right].
     \end{split}
 \end{equation}
\end{theorem}
From Theorem \ref{theorem reset hosidf} and \cite[Theorem 2]{10473227}, the open-loop HOSIDF of the system in Fig. \ref{fig:Block diagram cl} is as follows:
\begin{equation}
    \label{openloop HOSIDF}
    \mathcal{L}_n(\omega)=G(jn\omega)C_2(jn\omega)H_n(\omega)C_1(j\omega)e^{j(n-1)\angle C_1(j\omega)}.
\end{equation}

Despite LTI systems, there is no direct link between the open-loop and closed-loop frequency domain response for nonlinear controllers, especially in the case of the reset control system. Therefore, predicting the actual closed-loop performance is desirable using only frequency domain analysis. In Theorem \ref{Sn theorem}, a frequency domain-based performance prediction for RCSs is presented (\cite{saikumar2021loop}) under the following assumption.
\begin{assumption}
\label{Assumption two reset}
    We assume that only the first-order harmonic of $e_r$ contributes to resets, as in the closed-loop, it consists of a fundamental harmonic with a frequency of $\omega$, along with higher-order harmonics. Additionally, the RCS is input-to-state convergent (see \cite{hosseini2025frequency} for frequency domain stability analysis).
\end{assumption}
\begin{theorem}
    \label{Sn theorem}
    \cite[Theorem 4.1]{saikumar2021loop}
    Considering $r(t)=\sin{(\omega t)}$ and that Assumption \ref{Assumption two reset} holds, the closed-loop steady-state error $e_{ss}(t)$ can be written as 
    \begin{equation}
    \label{eq ess}
        e_{ss}(t)=\sum_{n=1}^{\infty}e_n(t),
    \end{equation}
    where
    \begin{equation}
    \label{eq en}
        e_n(t)=|S_n(\omega)|\sin{(n\omega t+\angle{S_n(\omega)})},
    \end{equation}
    with higher-order sensitivity function $S_n(\omega)$ as
    \begin{equation}
        \label{eq Sn}
        \resizebox{1\hsize}{!}{
        $S_n(\omega)=\begin{cases}
        \displaystyle
            \frac{1}{1+\mathcal{L}_1(\omega)},  \hspace{49.5mm} &\mathrm{for}\;\; n=1, \vspace{2mm}\\
            \displaystyle
            -\mathcal{L}_n(\omega)S_\mathrm{bl}(jn\omega)\left(|S_1(\omega)|e^{jn\angle{S_1{(\omega)}}}\right),  \hspace{11mm} &\mathrm{for}\;\;\mathrm{odd}\;\; n\geq2,\\
            0, \hspace{65.5mm} &\mathrm{for}\;\;\mathrm{even}\;\; n\geq2,
		\end{cases}$
  }
    \end{equation}
    where $S_\mathrm{bl}(jn\omega)=\frac{1}{1+L_\mathrm{bl}(jn\omega)}$ which $L_\mathrm{bl}(j\omega)$ is the base linear transfer function of the open-loop.
\end{theorem}
\section{Case study}
\label{sec: case study}
\begin{figure}
	\centering
 	\includegraphics[scale=0.30,trim=0 0 0 0,clip]{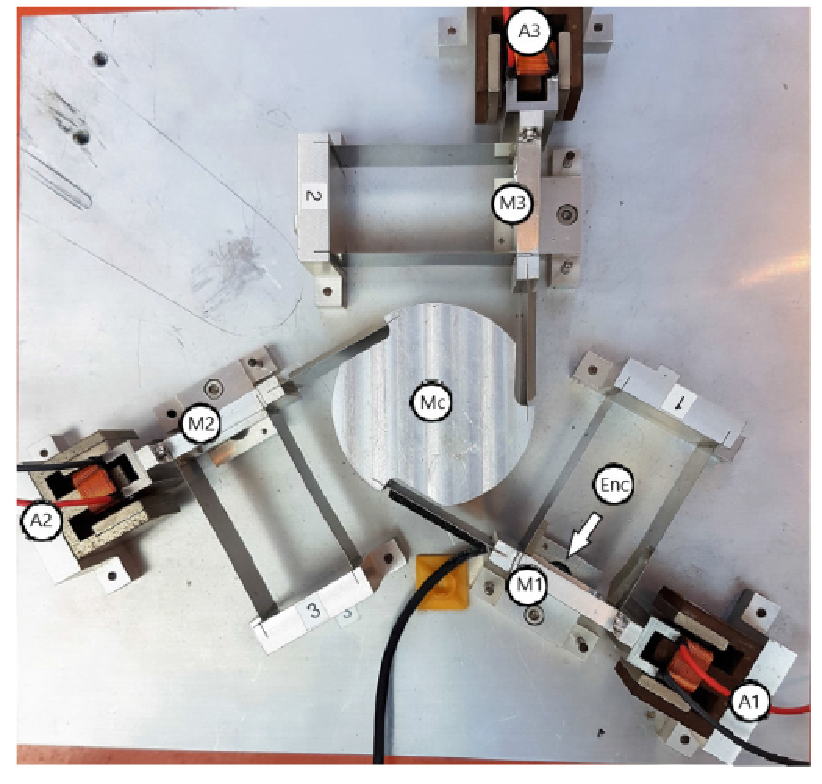}
	\caption{Planar precision positioning ‘Spider’ stage with voice coil actuator denoted
as A1, controlling the mass indicated as M1 and
constrained by leaf flexure.}
	\label{fig: spyder stage}
\end{figure}

In this paper, we address the vibration and disturbance rejection problem for a precision positioning stage (see Fig. \ref{fig: spyder stage}). The plant is modeled using its measured frequency response function, which can be approximated by the following transfer function:
\begin{equation}
\label{eq: Plant transfer function}
G(s)=\frac{9836 e^{-0.00027}}{s^2+8.737s+7376}.
\end{equation}
In this study, we use the above model of the actual system to be able to investigate its behavior under varying conditions, including the presence and absence of inputs such as noise and disturbances. A PID controller is designed to achieve a 30-degree phase margin while ensuring high low-frequency loop gain for improved disturbance rejection. However, due to time delay, the controller's ability to increase low-frequency gain is limited, as doing so requires either widening the bandwidth or raising the integral frequency, both of which reduce phase margin.

To overcome the mentioned limitations of PID controllers, we use the reset element to design a CgLp element (\cite{saikumar2019constant}) that provides phase lead around the frequency range of interest without affecting the gain behavior. We consider,
\begin{equation}
\label{eq CgLp(w)}
C_\mathrm{CgLp}= \mathcal{R}\, .\, C_\mathrm{lead},
\end{equation}
where
\begin{equation}
    \label{eq lead CgLp}
    C_\mathrm{lead}(s)=\frac{1+\frac{s}{\omega_r}}{1+\frac{s}{\omega_f}},
\end{equation}
with $\omega_f>>\omega_r$ (in this study $\omega_f=20\omega_r$).
The lead element ($C_\mathrm{lead}$) is placed after the reset component to prevent the amplification of noise in the reset action signal \( e_r(t) \) when the error signal \( e(t) \) is noisy. In many applications, the lead element is positioned either after the reset element or in parallel to it, helping to mitigate noise amplification in the reset signal \cite{10473227}.

The reset element $\mathcal{R}$ is designed such that its gain cancels the gain effect of the lead filter while providing less phase lag in its SIDF ($H_1(\omega)$) compared to the phase of $C_\mathrm{lead}^{-1}(s)$. Considering $\mathcal{R}$ as GFORE element ($A_r=-\omega_\alpha$, $B_r=1$, $C_r=\omega_\alpha$, $D_1=0$, $-1<A_\rho<1$) with the base linear transfer function 
\begin{equation}
    \label{eq: base linear GFORE}
    R(s)=\frac{1}{1+\frac{s}{\omega_{\alpha}}},
\end{equation}
where $\omega_{\alpha}$ is the corner frequency of the base linear transfer function of the GFORE element. To have the same magnitude of describing function as a linear low-pass filter $\left(1+\frac{s}{\omega_r}\right)^{-1}$ in both high frequencies ($\omega\rightarrow \infty$) and low frequencies ($\omega\rightarrow 0$), we set
\begin{equation}
    \label{eq corner frequency}
    \omega_\alpha=\frac{\omega_r}{\sqrt{1+\Theta_{\mathrm{D},\infty}^2}},
\end{equation}
where $\Theta_{D,\infty} := \lim_{\omega \to \infty} \Theta_\mathrm{D}(\omega)=\frac{4(1-A_\rho)}{\pi(1+A_\rho)}$. See \cite[Section 3]{van2024nonlinear} for more information.

\begin{figure}
	\centering
 	\includegraphics[scale=0.50,trim=0 8 0 0,clip]{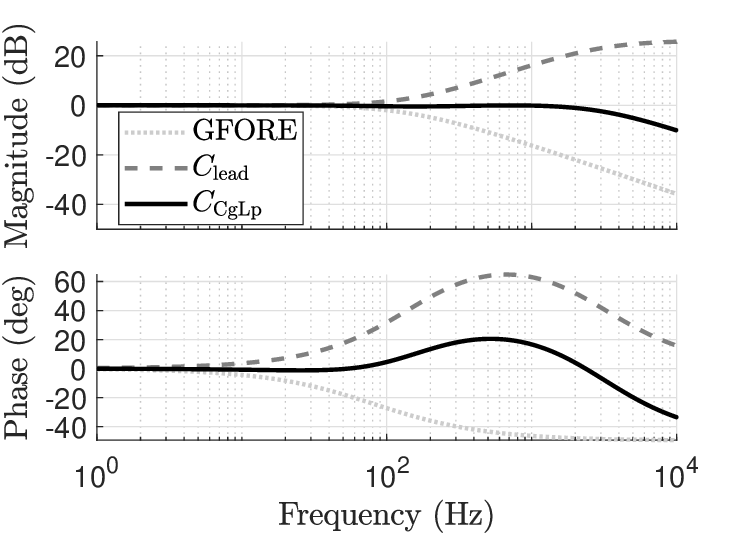}
	\caption{The SIDF of the CgLp element.}
	\label{fig: CgLp}
\end{figure}

\begin{figure}
	\centering
 	\includegraphics[scale=0.5,trim=0 0 0 0,clip]{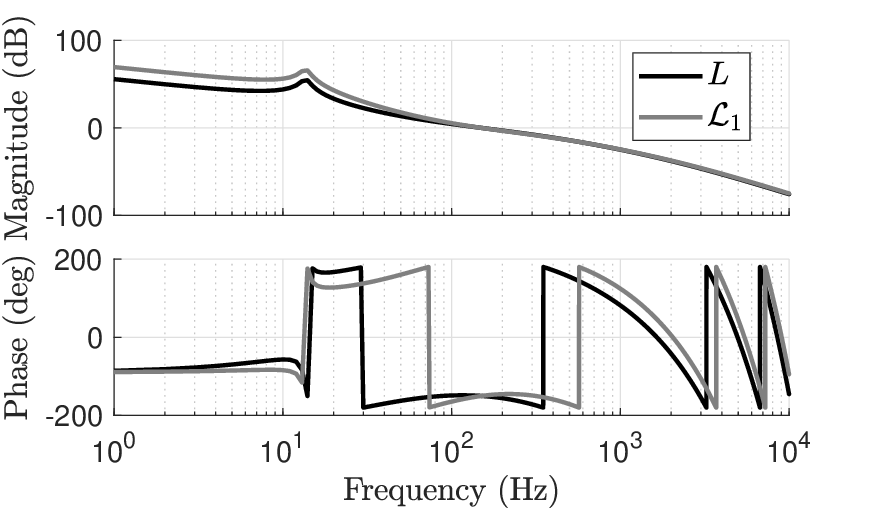}
	\caption{Open-loop frequency response of the plant with linear and nonlinear controller.}
	\label{fig: openloop}
\end{figure}

Having the plant in \eqref{eq: Plant transfer function}, we compare two controllers as 
\begin{equation}
\label{C-linear}
    C_\mathrm{L}=C_\mathrm{PID}\, .\, \frac{1}{1+\frac{s}{\omega_{f}}},
\end{equation}
and
\begin{equation}
    C_\mathrm{NL}=C_\mathrm{PID}\, .\, C_\mathrm{CgLp},
\end{equation}
where $C_\mathrm{PID}$ is defined as follows
\begin{equation}
	\label{PID1}
	C_{\text{PID}}(s)=k_p\bigg(1+\frac{\omega_i}{s}\bigg)\bigg(\frac{1+\frac{s}{\omega_d}}{1+\frac{s}{\omega_t}}\bigg),
\end{equation}
where $k_p \in \mathbb{R}$ represents the PID gain, ensuring zero dB gain at crossover frequency $\omega_c$ (defined as bandwidth), $\omega_i=\omega_c/10$ is the frequency at which integral action is stopped, differentiating action is started at $\omega_d=\omega_c/3$ and terminated at $\omega_t=3\omega_c$.
\begin{table}[]
\caption{Controller parameters}
\label{tab: CE1 parameters}
\resizebox{\columnwidth}{!}{%
\begin{tabular}{ccccccccc}

          & \large{$k_p$} &\large{$\omega_i \, [\mathrm{Hz}]$} & \large{$\omega_d \,[\mathrm{Hz}]$} & \large{$\omega_t \,[\mathrm{Hz}]$}& \large{$\omega_f \,[\mathrm{Hz}]$}& \large{$\omega_r \, [\mathrm{Hz}]$} &\large{$\omega_\alpha \,[\mathrm{Hz}]$}           &\large{$A_\rho$}    \\ \hline
\large{$C_{\mathrm{L}}$}       & \large{29.74}        &\large{$15$}& \large{$50$}                &\large{$450$}&\large{$3000$}& $-$ &$-$         & $- $ \\ 
\large{$C_{\mathrm{NL}}$}       & \large{29.85}        &\large{$50 $}& \large{$50$}                &\large{$450$}&\large{$3000$}&\large{ 150} &\large{$114.5$}         & \large{$0.2$} \\ \hline
\end{tabular}
}
\end{table}
In this work, we set the bandwidth as the maximum achievable one for the linear controller, \( C_{\mathrm{L}} \), to \(\omega_c = 150 \, \text{Hz}\) with a 30-degree phase margin. The corresponding PID controller parameters are derived using the rules-of-thumb method \cite{schmidt2020design}, as shown in Table. \ref{tab: CE1 parameters}. For the nonlinear controller, \( C_{\mathrm{NL}} \), we aim to achieve a 10-degree phase lead at the bandwidth frequency from the CgLp element, while maintaining only a 20-degree phase contribution from the PID controller. This configuration permits the use of a higher integral frequency, \(\omega_i=50\,\)Hz (instead of 15\,Hz in $C_\mathrm{L}$), which, in turn, results in a higher loop gain at lower frequencies. The PID and CgLp parameters for \( C_{\mathrm{NL}} \) are also provided in Table. \ref{tab: CE1 parameters}. The SIDF of \( C_{\mathrm{CgLp}} \) for the specified parameters is illustrated in Fig. \ref{fig: CgLp}. Notably, the gain of this filter remains near unity (0 dB) even one decade beyond the bandwidth frequency, while providing a positive phase at the bandwidth frequency. Fig. \ref{fig: openloop} illustrates the open-loop frequency response of the linear system as
\begin{equation}
    L(s)=C_\mathrm{L}(s)G(s),
\end{equation}
and the SIDF of the nonlinear control system as
\begin{equation}
    \label{eq: L nonlinear}
    \mathcal{L}_\mathrm{1}(\omega)=C_\mathrm{PID}(j\omega)C_\mathrm{lead}(j\omega)H_1(\omega)G(j\omega).
\end{equation}
It is evident that, in the case of nonlinear control, the incorporation of the reset element facilitates an increase in loop gain at low frequencies while maintaining the bandwidth and phase margin consistent with those of the linear control system.

However, thus far in this section, only the first-order SIDF characteristics of the reset element have been presented. In the subsequent sections, we will analyze the nonlinear control case based on the closed-loop higher-order SIDFs, aiming to minimize their impact on system performance.
\section{Shaping the closed-loop HOSIDFs}
\label{sec: shaping}
\begin{figure}
	\centering
 	\includegraphics[scale=0.55,trim=0 0 0 0,clip]{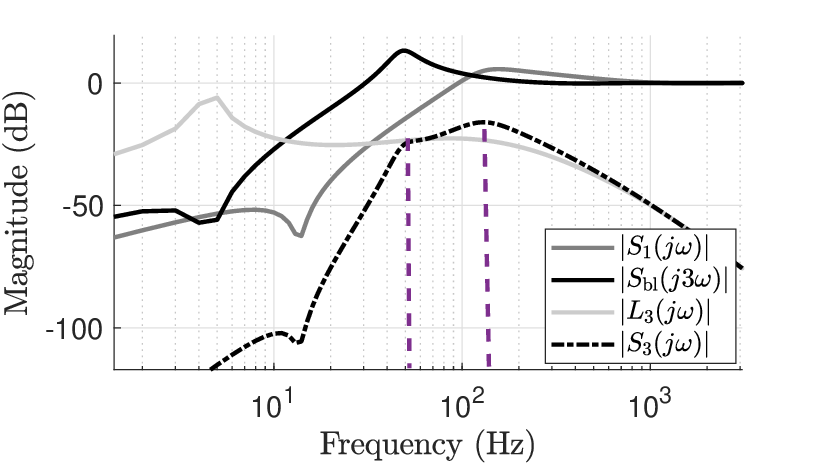}
	\caption{The third-order closed-loop sensitivity function and its components.}
	\label{fig: Sn wx}
\end{figure}

In this section, we analyze the closed-loop HOSIDFs of the designed nonlinear controller. According to Theorem \ref{Sn theorem}, the steady-state error comprises contributions from both the fundamental harmonic \( S_1(\omega) \) and higher-order harmonics \( S_n(\omega) \). Our focus is on \( S_3 \), as it exhibits the largest magnitude among the HOSIDFs. From \eqref{eq Sn}, we derive the following relationship:
\begin{equation}
\label{eq: |S3|}
|S_3(\omega)| = |\mathcal{L}_3(\omega)| |S_\mathrm{bl}(j3\omega)| |S_1(\omega)|,
\end{equation}
where \( |S_3(\omega)| \) and its components are illustrated in Fig. \ref{fig: Sn wx}, assuming \( C_1 = 1 \) and \( C_2 = C_\mathrm{lead} \cdot C_\mathrm{PID} \).

It is evident that \( \max{(|S_1(\omega)|)} \) occurs at or near the bandwidth frequency \( \omega_c \), while the peak of \( |S_\mathrm{bl}(j3\omega)| \) appears around \( \omega_c / 3 \). Consequently, the third-order sensitivity function \( |S_3(\omega)| \) has a relatively large magnitude around these peaks, which is not desired. To effectively reduce \( |S_3(\omega)| \) in these regions, it is crucial to shape \( |\mathcal{L}_3(\omega)| \) while preserving \( |S_\mathrm{bl}(j3\omega)| \) and \( |S_1(\omega)| \) unchanged.




 
 Based on existing literature, shaping open-loop HOSIDFs while preserving the first-order harmonic characteristics unchanged can be achieved in two ways: first, by altering the sequence of elements within the loop \cite{cai2020optimal}; and second, by applying pre- and post-filtering to the reset element using two filters that are inverses of each other \cite{karbasizadeh2022continuous}. In this work, we propose a remedy to address the trade-off in the sequencing of CgLp elements to target HOSIDFs around the bandwidth frequency $\omega_c$. Subsequently, we introduce a novel filtering method designed to specifically target HOSIDFs around $\omega_c/3$.

\subsection{Tunable Lead CgLp Element}\label{sec: Tunable}
Regarding the equation in \eqref{openloop HOSIDF}, the HOSIDFs of the open-loop ($\mathcal{L}_n(\omega)$) can be varied by changing the sequence of $C_1$ and $C_2$ in the loop. The effect of element sequencing on the open-loop HOSIDFs was first examined in \cite{cai2020optimal}. Based on the findings in \cite{cai2020optimal}, minimizing open-loop nonlinearity ($\mathcal{L}_n(j\omega)$) requires positioning the lead action before the reset element in the control loop. Consequently, for the CgLp element, the $C_\mathrm{lead}$ filter should precede the GFORE element. However, as we explained in the previous section, having lead action before the reset element can easily magnify the noise in error $e(t)$ and transfer it to the reset signal $e_r(t)$ and cause many unwanted reset actions. On the other hand, we still want to use the benefit of having lead action before the reset element to reduce the higher-order harmonics based on the result in \cite{cai2020optimal}. Thus, we introduce the tunable lead CgLp element as illustrated in Fig. \ref{fig: Tunable lead CgLp}, where \( C_\mathrm{lead} = L_1 \, . \, L_2 \), with
\begin{equation}
    \label{eq L1}
    L_1(s)=\frac{1+\frac{s}{\omega_r}}{1+\frac{s}{\omega_x}},
\end{equation}
and
\begin{equation}
    \label{eq L2}
    L_2(s)=\frac{1+\frac{s}{\omega_x}}{1+\frac{s}{\omega_f}},
\end{equation}
where \( \omega_r \leq \omega_x \leq \omega_f \).
\usetikzlibrary {arrows.meta}
\tikzstyle{sum} = [draw, fill=white, circle, minimum height=0.0em, minimum width=0.0em, anchor=center, inner sep=0pt]
\tikzstyle{block} = [draw,thick, fill=white, rectangle, minimum height=2.5em, minimum width=3em, anchor=center]
\begin{figure}[!t]
	\centering
	\begin{scaletikzpicturetowidth}{0.8\linewidth}
		\begin{tikzpicture}[scale=\tikzscale]
			\node[coordinate](input) at (0,0) {};
			
			\node[sum] (sum3) at (6.3,0) {};
			

			\node[block] (lead) at (3.2,0) {$\mathcal{R}$};
			\node[block] (controller) at (4.9,0) {$L_2$};
			\node[block] (fo-higs) at (1.5,0) {$L_1$};
			
			\node[coordinate](output) at (7.2,0) {};

			\draw[arrows = {-Latex[width=8pt, length=10pt]}] (input)  -- node[above]{} (fo-higs);
			\draw[arrows = {-Latex[width=8pt, length=10pt]}] (fo-higs) --node [above]{} (lead);
			\draw[arrows = {-Latex[width=8pt, length=10pt]}] (lead)  --node[above] {} (controller);
			\draw[arrows = {-Latex[width=8pt, length=10pt]}] (controller)  --node[above] {} (sum3);

              \draw [color=gray,thick,dashed](0.3,-0.75) rectangle (5.7,0.75);
			\node at (0.3,1) [left,color=gray] {$\text{CgLp}$};
			
		\end{tikzpicture}
	\end{scaletikzpicturetowidth}
	\caption{Tunable lead CgLp.}
	\label{fig: Tunable lead CgLp}	
\end{figure}
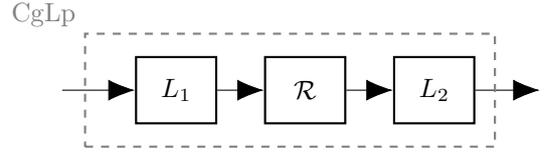
With this architecture, we can adjust the placement of a partial lead component relative to the reset element, to achieve an optimal balance between noise amplification and the reduction of higher-order harmonics. Since \( \omega_r \) is established at the bandwidth frequency, and given that \( \omega_r \leq \omega_x \leq \omega_f \), we anticipate improvements in the HOSIDFs around or beyond the bandwidth frequency. To determine the optimal value for \(\omega_x\), we vary \(\omega_x\) within the range of \([\omega_r, \omega_f]\). This approach allows us to identify the \(\omega_x\) value that minimizes error, depending on the system’s signal-to-noise ratio (SNR), by balancing noise reduction with suppression of higher-order harmonics. This process is detailed in the Section. \ref{sec: Frequency Domain Validation}.

However, since the CgLp is primarily active around the bandwidth frequency (here, \(\omega_r = \omega_c\)), this tunable lead method is not effective in adequately targeting the HOSIDFs below \(\omega_c\). Additionally, as the peak of \(|S_\mathrm{bl}(j3\omega)|\) leads to an increase in \(|S_3(\omega)|\) around \(\omega_c/3\), the next section discusses an approach to target these frequencies separately.

\subsection{Filtered CgLp Element}
In \cite{karbasizadeh2022continuous}, a novel reset control system architecture is proposed, employing a lead filter before, and a lag filter after the reset element to mitigate nonlinearity, forming the "continuous reset element." However, to limit noise amplification in the reset signal, this study seeks to minimize the use of lead filters before the reset element. Thus, we directly examine the closed-loop HOSIDFs and analyze the impact of filtering on them.

Consider the closed-loop block diagram in Fig. \ref{fig:Block diagram cl}, where a filter \( N \) is implemented in \( C_1 \), and the inverse of this exact filter, \( N^{-1} \), is implemented in \( C_2 \). Accordingly, if we denote \( S_n^{'} \) as the higher-order sensitivity function in presence of \( N \) and \( N^{-1} \), we have:
\begin{equation}
    \frac{S_n^{'}(\omega)}{S_n(\omega)} = \frac{-\mathcal{L}_n^{'}(\omega) S_\mathrm{bl}^{'}(jn\omega) |S_1^{'}(\omega)| e^{jn \angle{S_1^{'}(\omega)}}}{-\mathcal{L}_n(\omega) S_\mathrm{bl}(jn\omega) |S_1(\omega)| e^{jn \angle{S_1(\omega)}}},
\end{equation}
where 
\begin{equation}
    \mathcal{L}_n^{'}(\omega) = G(jn\omega) C_2^{'}(jn\omega) H_n(\omega) C_1^{'}(j\omega) e^{j(n-1)\angle C_1^{'}(j\omega)}.
\end{equation}
With \( C_1^{'}(j\omega) = C_1(j\omega) N(j\omega) \) and \( C_2^{'}(j\omega) = C_2(j\omega)  \linebreak N^{-1}(j\omega) \), we arrive at
\begin{equation}
\label{eq: NN-1}
    \frac{\left|S_n^{'}(\omega)\right|}{\left|S_n(\omega)\right|} = \left|N(j\omega)\right| \left|N^{-1}(jn\omega)\right|.
\end{equation}
Thus, if the filter \( N \) can be designed such that around the frequencies of interest \( \left|N(j\omega)\right| \left|N^{-1}(jn\omega)\right| < 1 \), then the magnitude of \( S_n^{'}(\omega) \) can be reduced within that range.
\section{Illustrative Example}
\label{Sec: validation}
Here, we address a disturbance rejection problem, focusing on minimizing the error caused by a 40 Hz sinusoidal disturbance, \( d = 0.25 \sin{(2\pi \cdot 40 \, t)} \), in the presence of white noise, while assuming no error arises from the reference \( r(t) \) due to an ideal feedforward controller ensuring perfect tracking. As shown in Fig. \ref{fig: openloop}, \( C_\mathrm{NL} \) is expected to outperform \( C_\mathrm{L} \) in disturbance suppression due to its higher gain at the disturbance frequency. However, Fig. \ref{fig: Sn wx} reveals that \( C_\mathrm{NL} \) introduces additional error due to its nonlinear characteristics. Our objective is to minimize this error using two proposed methods: the tunable lead \( \text{CgLp} \) and the filtered \( \text{CgLp} \).

\subsection{Frequency Domain Validation}
\label{sec: Frequency Domain Validation}
To tune \(\omega_x\) in \eqref{eq L1} and \eqref{eq L2}, the signal-to-noise ratio must be known. For an SNR of 47.1 dB, \(\omega_x = 360\)\,Hz minimizes the root mean square (RMS) error compared to \(\omega_x = \omega_r\) and \(\omega_x = \omega_f\). See Appendix \ref{app: wx optimization} for details on \(\omega_x\) tuning.

\begin{figure}
	\centering
\includegraphics[scale=0.50,trim=0 0 0 0,clip]{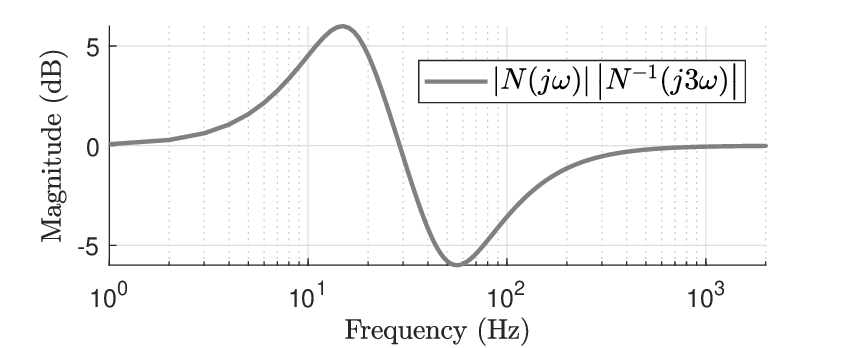}
	\caption{The $\frac{\left|S_3^{'}(\omega)\right|}{\left|S_3(\omega)\right|}$ ratio.}
	\label{fig: Notch-Anotch}
\end{figure}

To design the filter \( N \), we aim to attenuate the magnitude of \( S_3(\omega) \) around $\omega_c/3=50$ Hz, where it exhibits a relatively high peak, which is related to the peak of \( |S_\mathrm{bl}(j3\omega)| \). Since \( N \) is placed before the reset element and \( N^{-1} \) is placed after it, it is essential to ensure that \( |N(j\omega)| \leq 1 \) for all \( \omega \in \mathbb{R} \) to prevent any noise amplification. Therefore, we define \( N \) as a notch filter, expressed as follows:
\begin{equation}
    \label{eq: notch}
    N(s)= \frac{{s^2}/{\omega_n^2} + {s}/({Q_1 \omega_n}) + 1}{{s^2}/{\omega_n^2} + {s}/({Q_2 \omega_n}) + 1},
\end{equation}
where \( \omega_n, Q_1, Q_2 \in \mathbb{R} \). At the target frequency $\omega_n$ we have:
\begin{equation}
    \label{eq: N(jw)}
    |N(j\omega)|_{\omega=\omega_n}=\frac{Q_2}{Q_1},
\end{equation}
where from \eqref{eq: NN-1}, approximately we have $\frac{\left|S_3^{'}(\omega)\right|_{\omega=\omega_n}}{\left|S_3(\omega)\right|_{\omega=\omega_n}}\approx\frac{Q_2}{Q_1}$. In this study, we selected $\omega_n = 50$ Hz, $Q_1 = 1$, and $Q_2 = 0.4$ to reduce $\left|S_3(\omega)\right|$ by more than half at 50 Hz. Fig. \ref{fig: Notch-Anotch} shows the plot of  \(
\left|N(j\omega)\right|\left|N^{-1}(j3\omega)\right|\) where the filter is able to effectively reduce $\left|S_3(\omega)\right|$ at the target frequency, with an amplitude increase at lower frequencies. This increase is negligible, as $S_1$ and $S_3$ have very small magnitudes in this range, making further harmonic suppression unnecessary.

\begin{figure}
	\centering
 	\includegraphics[scale=0.4,trim=0 0 0 0,clip]{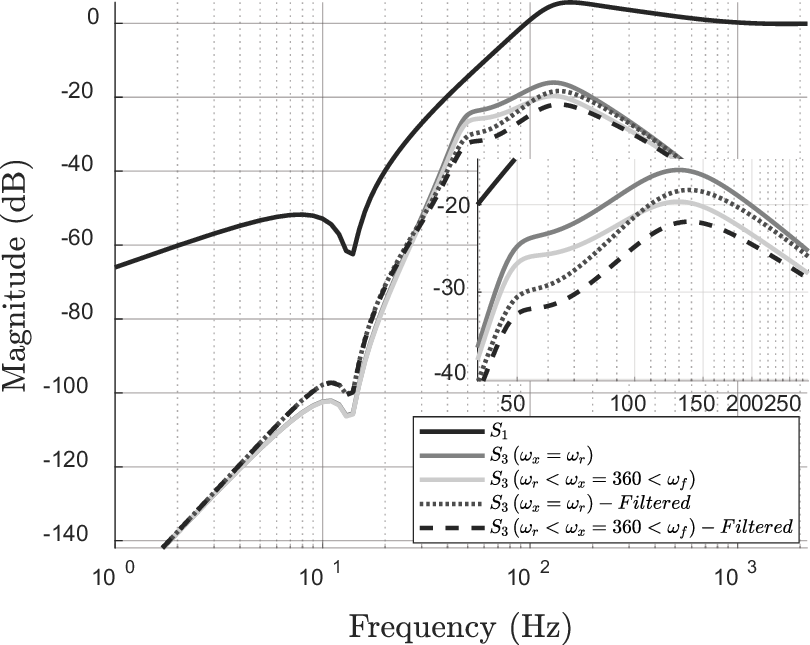}
	\caption{The first- and third-order closed-loop SIDFs for normal, tunable lead, and filtered tunable lead CgLp.}
	\label{fig: S3 all}
\end{figure}

Having designed both the Tunable Lead, and Filtered CgLp, we calculate \( |S_3(\omega)| \) for \( C_1 = L_1 \, . \, N \) and \( C_2 = L_2 \, . \, N^{-1} \, . \, C_{\text{PID}} \), comparing it with the case in Fig. \ref{fig: Sn wx}. As shown in Fig. \ref{fig: S3 all}, the Tunable Lead architecture reduces the magnitude of \( S_3 \) around the bandwidth (150 Hz). Additionally, applying the notch filter and its inverse before and after the GFORE, effectively attenuates the peak of \( |S_3(\omega)| \) near one-third of the bandwidth (50 Hz).
\subsection{Time Domain Validation}
\label{sec: time domain validation}
\begin{figure}
	\centering
\includegraphics[scale=0.5,trim=0 0 0 0,clip]{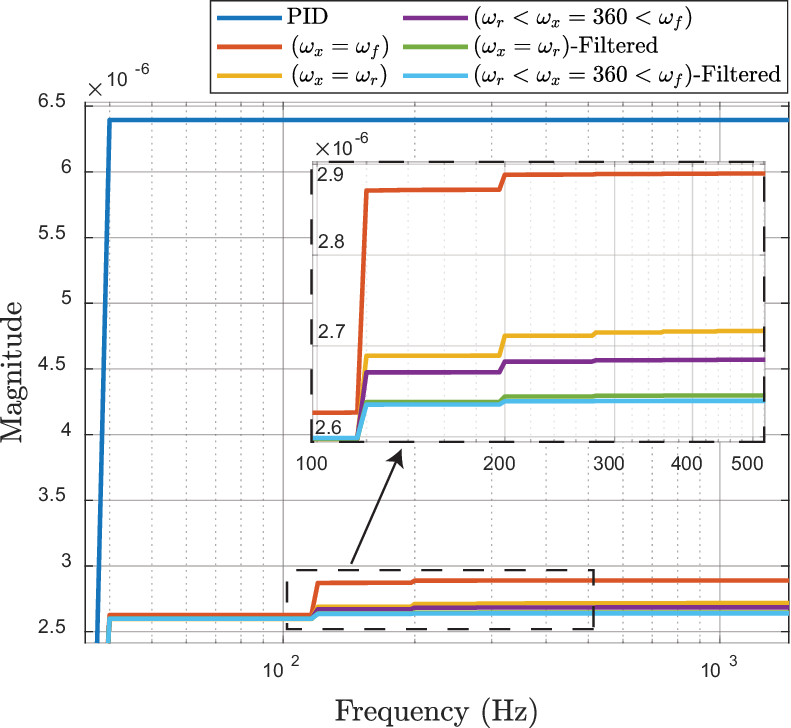}
	\caption{The CPSD of the errors.}
	\label{fig: CPSD all}
\end{figure}

We implement the linear controller (\(C_\mathrm{L}\)) from \eqref{C-linear} and four CgLp cases, whose closed-loop frequency responses are shown in Fig. \ref{fig: S3 all}. As mentioned, we consider a zero-reference tracking case with a disturbance \( d = 0.25 \sin{(2\pi \cdot 40 \, t)} \) and white noise.
Fig. \ref{fig: CPSD all} presents the cumulative power spectral density (CPSD) of the error for each controller, measured after allowing sufficient time for transient behavior to fully dampen.
The improvement of the nonlinear controller over the linear controller is evident from these results. Specifically, the peak of the CPSD of the error, proportional to the RMS value, is reduced for all nonlinear controllers. Among them, the \( C_\mathrm{lead} \)-GFORE case (\( \omega_x = \omega_f \)) shows the highest error. This aligns with prior findings that applying lead action before the reset element in the presence of noise leads to a noisy reset trigger signal, \( e_r \). However, it is also evident that in the GFORE-\( C_\mathrm{lead} \) configuration (\( \omega_x = \omega_r \)), further refinement is necessary to mitigate the nonlinearity introduced by the controller. Consequently, cases that incorporate the tunable lead \( C_\mathrm{gLp} \), the filtered \( C_\mathrm{gLp} \), or a combination of both show a marked reduction in higher-order harmonics. These results align with the frequency domain responses of the third-order closed-loop SIDFs across the different controllers, as they share the same magnitude of \( |S_1(\omega)| \) but exhibit distinct \( |S_3(\omega)| \).
\section{Conclusion}
In conclusion, this paper has addressed the challenges posed by nonlinearity in RCSs and presented methods to mitigate these effects within the context of the CgLp filter. By decomposing the lead action in the CgLp filter into two components and introducing an additional parameter, this work offers a novel approach to balancing the trade-offs between noise amplification and nonlinearity introduced by reset elements. Additionally, a filtering technique has been proposed that targets specific frequencies, providing a promising solution for reducing nonlinearity in RCSs, especially in the presence of noise. The simulation results in both the frequency and time domains validate the effectiveness of these methods, highlighting their potential to enhance the performance of RCSs in practical applications.
A more sophisticated tuning design for the notch filter in this study, incorporating all the HOSIDFs rather than just the third, would further enhance the performance of RCSs.\cite{Abl:56,AbTaRu:54,Keo:58,Pow:85}

\bibliography{ifacconf}

\appendix
\section{Optimal value for}
\label{app: wx optimization}
\begin{figure}
	\centering
 \includegraphics[scale=0.5,trim=0 0 0 0,clip]{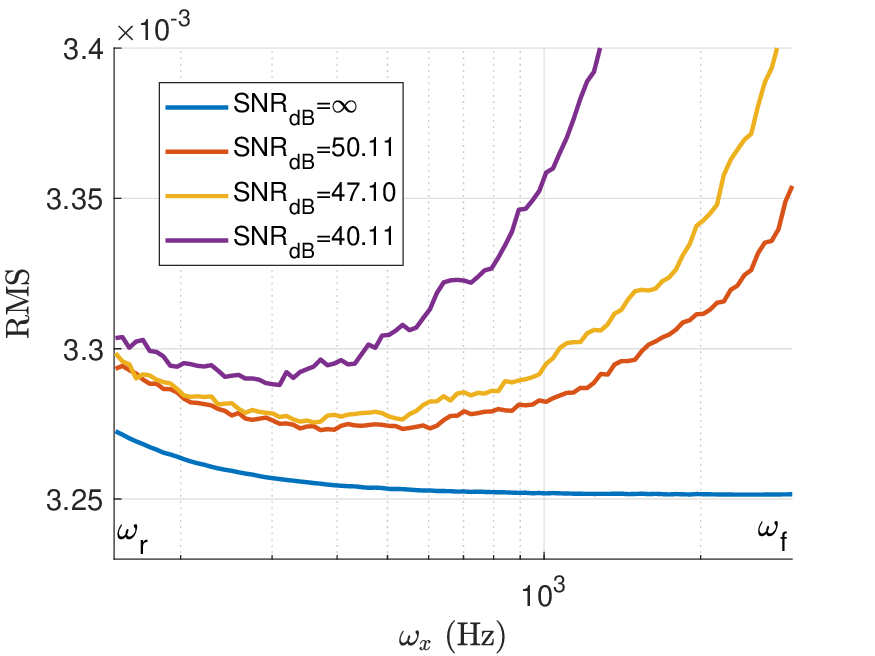}
	\caption{The RMS value for the cases where different noises are considered for tunable lead CgLp controller.}
	\label{fig: SNR}
\end{figure}

Consider the closed-loop system shown in Fig. \ref{fig:Block diagram cl}, with \( C_1 = L_1 \) and \( C_2 = L_2 \, . \, C_\mathrm{PID} \), and corresponding parameters listed in Table \ref{tab: CE1 parameters}. For the disturbance rejection scenario discussed in Section \ref{sec: time domain validation}, where \( r(t) = 0 \) and \( d = 0.25 \sin{(2\pi \cdot 40 \, t)} \), we analyze the system's response in the presence of white noises with different power levels, resulting in different SNRs. In Fig. \ref{fig: SNR}, by varying \( \omega_x \) within the range of \( [\omega_r, \omega_f] \), it is observed that there exists a specific value \( \omega_r \leq \omega_x \leq \omega_f \) that minimizes the root mean square (RMS) error in the system. Note that previously, in the presence of noise, the \( \text{CgLp} \) was implemented using the \(\text{GFORE-}C_{\text{lead}}\) structure. However, as shown in Fig. \ref{fig: SNR}, there exists a point where this new approach results in reduced error compared to the original \(\text{GFORE-}C_{\text{lead}}\) implementation.

\end{document}